\documentclass[aps,showpacs,preprintnumbers,amsmath,amssymb,superscriptaddress,floatfix,a4paper,nofootinbib,11pt,floatfix,nofootinbib,onecolumn]{revtex4}
\usepackage{graphicx}
\usepackage{bm}
\usepackage{rotating}
\usepackage{array}
\usepackage{xcolor}
\usepackage{amsmath,amssymb}
\usepackage{mathrsfs}
\usepackage{graphicx}
\usepackage{color}
\usepackage{subfigure}
\usepackage{fancyhdr}
\usepackage{multirow}
\usepackage{float}
\usepackage{epsfig}
\usepackage{amsfonts}
\usepackage{bm}

\newcommand{\ba}{\begin{eqnarray}}
\newcommand{\ea}{\end{eqnarray}}

\begin{document}

\title{\Large \bf Exact Solutions of (deformed) Jackiw-Teitelboim Gravity}

\author{Davood Momeni} 
\email{davood@squ.edu.om}
\affiliation{Department of Physics, College of Science, Sultan Qaboos University, P.O. Box 36, \\Al-Khodh 123, Muscat, Sultanate of Oman}

\author{Phongpichit Channuie} 
\email{channuie@gmail.com}
\affiliation{College of Graduate Studies, Walailak University, Thasala, \\Nakhon Si Thammarat, 80160, Thailand}
\affiliation{School of Science, Walailak University, Thasala, \\Nakhon Si Thammarat, 80160, Thailand}
\affiliation{Research Group in Applied, Computational and Theoretical Science (ACTS), \\Walailak University, Thasala, Nakhon Si Thammarat, 80160, Thailand}

\begin{abstract}
It is well known that Jackiw-Teitelboim (JT) gravity posses the simplest theory on 2-dimensional gravity. The model has been fruitfully studied in recent years. In the present work, we investigate exact solutions for both JT and deformed JT gravity recently proposed in the literature. We revisit exact  Euclidean solutions for Jackiw-Teitelboim gravity using all the non-zero components of the dilatonic equations of motion using proper integral transformation over Euclidean time coordinate. More precisely, we study exact solutions for hyperbolic coverage, cusp geometry and another compact sector of the AdS$_2$ spacetime manifold. 
\end{abstract}


\maketitle
\date{\today}

\section{Introduction}

The discovery of exactly solvable models of quantum gravity, starting with Kitaev \cite{Kitaev2015} and subsequent investigations \cite{Polchinski:2016xgd,Maldacena:2016hyu} posses one of the most exciting developments over the past few years. Since then, finding other models solvable to the same extent would be highly valuable, in particular to test the robustness of the ideas. Especially, in holographic duality, we describe a boundary theory on a manifold $\cal M$ via a bulk description on a manifold $\cal N$ whose boundary is governed by $\cal M$. The authors of Ref.\cite{Saad:2019lba} demonstrated that the AdS  bulk theory is dual to Hermitian random matrix theory. It provides a concrete and explicit example for lower dimensional gauge/gravity dualities beyond the standard AdS/CFT. Basically here the boundary theory is a nearly CFT and conforaml invariance and the dual bulk theory is UV free model for AdS gravity. Its correlation functions in various incarnations were later studied by many authors, see for example \cite{Bagrets:2016cdf,Stanford:2017thb,Mertens:2017mtv,Lam:2018pvp,Mertens:2018fds,Blommaert:2018oro,Kitaev:2018wpr,Yang:2018gdb,Iliesiu:2019xuh} including higher genus and random matrix descriptions \cite{Saad:2019lba}. 
Concretely, JT gravity is a simple model of a real scalar field $\phi$ coupled to gravity in two dimensions \cite{Jackiw:1984je,Teitelboim:1983ux}. Pure JT gravity is the limiting case $D\to 2$ of a typical classical general relativity (GR) obtained after applying a suitable {\it tricky} conformal transformations \cite{Mann:1992ar}. One starts from a general $D$-dimensional Einstein Hilbert-action including arbitrary $D$-dimensional matter action $\mathcal{L}_M$ given by 
\begin{eqnarray}
S=-\frac{1}{\kappa_D}\int d^Dx\sqrt{-g}R+\int d^Dx \sqrt{-g}\mathcal{L}_M\,,
\end{eqnarray}  
where $\kappa_D$ denotes a $D$-dimensional gravitational coupling. It was found that when taking $D\to 2$ applied on both an action and a coupling,  $\lim_{D\to 2} \kappa_D\sim \mathcal{O}(1-\frac{D}{2})$ and $\lim_{D\to 2} G_{\mu\nu}\sim \mathcal{O}(1-\frac{D}{2})$. We then subtract a term (total derivative) from the action in the limiting regime. After applying a conformal transformation on the second subtracted term and taking the limit $D\to 2$, we end up with a simple scalar field theory where the scalar field (dilaton) coupled to the scalar curvature analogous to $\phi R$. There is a possibility to redefine the gravitational coupling in this limiting case to remove any ambiguity in the theory. Furthermore, JT gravity action can be obtained from a suitable dimensional reduction of a general four-dimensional GR when the spacetime is time independent, spherically symmetric. The resulting theory is a subclass of the dilaton theories as an attempt to define effective theory for the gravity at large distances \cite{Grumiller:2010bz} .

For the case of negative cosmological constant where, the bulk action in Euclidean signature is governed by 
\begin{eqnarray}
S=-\frac{1}{2}\int_{\cal N} d^2x\sqrt{g}\phi(R+2) + ...\,,
\end{eqnarray}
where ellipses denote a topological term, e.g. Euler characteristic, $R$ is the scalar curvature of the metric tensor $g$. Evidently, the model has been fruitfully studied in recent years. These include for example the works done by Refs.\cite{Almheiri:2014cka,Maldacena:2016upp,Engelsoy:2016xyb,Harlow:2018tqv,Witten:2020wvy,Witten:2020ert,Sarosi:2017ykf}. The solutions within JT gravity in the presence of nontrivial couplings between the dilaton and the Abelian $1$-form were constructed in \cite{Lala:2019inz}. As an attempt to study viable deformations of the original JT gravity, Maxfield and Turiaci in Ref.\cite{Maxfield:2020ale}, showed that JT gravity can be considered as a near extremal of the AdS 3d gravity via path integral. Also in another recent paper \cite{Turiaci:2020fjj}, the authors demonstrated that the deformed JT gravity action can be derived from the the minimal string theory. Later, Witten in Ref.\cite{Witten:2020wvy} found  that a simple correspondence of JT gravity with a random matrix is possible in part since JT gravity is simple. Moreover, the results of Ref.\cite{Saad:2019lba} showed that JT gravity in two dimensions is dual not to any particular quantum system but to a random ensemble of quantum systems. Additionally, deformed JT gravity has been investigated in \cite{Witten:2020ert} by adding a self-coupling of the scalar field to the original JT gravity action. Witten showed that the resulting model is still dual to a random ensemble of quantum systems having a different density of energy levels.

To generalize JT gravity to a large class of models, let us emphasize a particular example of a self-coupling of the scalar field proposed in Refs. \cite{Maxfield:2020ale},\cite{Witten:2020ert}. Consider an action for a scalar field $\phi$ and a metric $g$ with each term having at most two derivatives. The most general possible action may take the form 
\begin{eqnarray}
S=-\frac{1}{2}\int_{\cal N} d^2x\sqrt{g}\Big(F(\phi)R + G(\phi)|\nabla \phi|^{2} + V(\phi)\Big)\,,\label{mod2D}
\end{eqnarray}
where $F(\phi),\,G(\phi)$ and $V(\phi)$ are functions depending on the (dilaton) field $\phi$. It was noticed that quantum aspects in the theory governed by the action (\ref{mod2D}) have been studied in detail in Ref.\cite{Nojiri:2000ja}. However, the above action can be simplified using a Weyl transformation of the metric together with a reparametrization of the scalar field $\phi$. With this transformation, we find that two of the three functions can be eliminated. Hence we can end up with the case of just single function: 
\begin{eqnarray}
S=-\frac{1}{2}\int_{\cal N} d^2x\sqrt{g}\Big(\phi R + U(\phi)\Big)\,.\label{ddJT}
\end{eqnarray}
In order to recover the result present in \cite{Witten:2020ert}, we can introduce a function $U(\phi)=2\phi + W(\phi)$ to Eq.(\ref{ddJT}), where $W(\phi)$ denotes the departure from original JT gravity. We will investigate the solutions for deformed JT gravity in Sec.\ref{dJT}. However, adding a self-coupling of the scalar field is not the only possibility to deform JT gravity.

In the present work, we investigate exact solutions for both JT and deformed JT gravity recently proposed in the literature. We revisit exact Euclidean solutions for Jackiw-Teitelboim gravity using all the non-zero components of the dilatonic equations of motion using proper integral transformation over Euclidean time coordinate in Sec.\ref{JT}. More precisely, in Sec.\ref{sec3}, we study exact solutions for hyperbolic coverage, cusp geometry and another compact sector of the AdS$_2$ spacetime manifold. In addition, we study dJT gravity and examine its solution in Sec.\ref{dJT}. We finally conclude our findings in the last section.

\section{Jackiw-Teitelboim gravity}
\label{JT}
Let us analyze the equations of motion of the Jackiw-Teitelboim theory. We now carry out corresponding calculations in JT gravity by starting with the JT action of the form \cite{Maldacena:2016upp}
\begin{eqnarray}
S=-\frac{1}{2}\Big[\int \sqrt{g}\phi(R+2)d^2x+2\int_{\partial}\phi_{\rm bdy}K\Big],\,\label{S1}
\end{eqnarray}
where $g,\,R,\,K,\,\phi$, and $\phi_{\rm bdy}$ refer to the metric, scalar curvature,
extrinsic curvature, dilaton field and its value on the boundary, respectively. Note here that we are working with units $\kappa^2=1,\,L_{{\rm AdS}_{2}}\,\equiv 1$. Using the action (\ref{S1}), Einstein field equations can be directly derived to obtain
\begin{eqnarray}
R+2&=&0\label{eom-g},\\ \nabla_{\mu}\nabla_{\nu}\phi-g_{\mu\nu}\nabla_{\alpha}\nabla^{\alpha}\phi+g_{\mu\nu}\phi&=&0\label{eom-phi}.
\end{eqnarray}
Note that in two dimensions, Einstein tensor follows $G_{\mu\nu}\equiv 0$. The second EoM (\ref{eom-phi}) gives the energy-momentum tensor $T_{\mu\nu}=0$. The first equation (\ref{eom-g}) gives us the scalar curvature for pure AdS$_2$. However it cannot fix any type of the geometry. Basically, we are working in the probe limit because the dilaton field doesn't backreact on the geometry. As a result, we can use any type of the AdS$_2$ metric and invoke an Euclidean form in the coordinates $(t,z)$, where $z=0$ is an AdS boundary. Because of the linear coupling to the dilaton field, the metric in the bulk is localized to AdS$_2$ so that we can write
\begin{eqnarray}
ds^2=\frac{dt^2+dz^2}{z^2}\label{metric1}.
\end{eqnarray}
Therefore, the non-vanishing components of Eq.(\ref{eom-phi}) for $(tt),\,(tz),\,(zz)$ for a dilaton profile $\phi=\phi(t,z)$ read
\begin{eqnarray}
&&(tt): \phi''-\frac{\phi}{z^2}=0\label{tt}\\&&
(tz): \dot{\phi}'+\frac{\dot{\phi}}{z}-\ddot{\phi}-\phi''+\frac{\phi}{z^2}=0\label{tz}\\&&
(zz): \ddot{\phi}-\frac{\phi}{z^2}=0\,, \label{zz}
\end{eqnarray}
where \lq\lq primes\rq\rq\, represent derivative with respect to $z$, and \lq\lq dots\rq\rq\, denote derivative with respect to $t$. In Ref.\cite{Maldacena:2016upp}, one solution of the above system takes the form 
\begin{eqnarray}
\phi(t,z)=\frac{\alpha+\gamma t+\delta(t^2+z^2)}{z}\,.
\end{eqnarray}
The authors used this particular solution to study boundary term in particular to define boundary field value $\phi_{\rm bdy}=\phi(t,\epsilon)|_{\epsilon\to 0}$. However, we are going to quantify a more general solution. Here we will show that the above solution is just a particular solution and it is a member of a more general family of the exact solutions. In addition to the system of the PDEs given in Eqs.(\ref{tt}-\ref{zz}), we have a trace equation of the EoM (\ref{eom-phi}) as a simple Helmholtz equation on AdS$_2$,
\begin{eqnarray}
&&\nabla_{\alpha}\nabla^{\alpha}\phi-2\phi=0\label{eom-trace}.
\end{eqnarray}
If we rewrite the above trace equation in the background (\ref{metric1}) we simply obtain
\begin{eqnarray}
&&\ddot{\phi}+\phi''-2\frac{\phi}{z^2}=0\label{trace}.
\end{eqnarray}
We will obtain eq. (\ref{trace}) just by adding Eq.(\ref{tt}) and Eq.(\ref{zz}) together. Let us do the following: We add the unused equation (\ref{tz}) to the trace equation (\ref{trace}), we find the following PDE :
\begin{eqnarray}
&&z(z\dot{\phi})'-\phi=0\label{eqphi}.
\end{eqnarray}
Any solution to the above PDE can be used to solve the trace equation (\ref{trace}) and consequently to solve eqs. (\ref{tt}),(\ref{zz}) as well as solve the off diagonal equation (\ref{tz}). By the above consistency to find the scalar field profile we only need to solve a single linear PDS given in (\ref{eqphi}). We will find an exact dilaton profile $\phi(t,z)$. Particularly, we want to quantify $\phi_{\rm bdy}=\phi(t,z=0)$. 

\section{Exact solutions}
\label{sec3}
In the present section, we quantify all possible solutions for Eqs.(\ref{tt}-\ref{zz}) and (\ref{trace}) by considering the hyperbolic coverage, cusp geometry and another compact sector of the AdS$_2$ spacetime manifold. Before going into details of determining the solutions. Let us clarify the equations and the solvability of the other equations. In JT gravity after fixing the AdS metric, the only remaining equation is the scalar field equation written in the Eq.(\ref{eom-phi}) of the current version. This scalar field equation is derived by varying the action of JT gravity with respect to the metric. In two dimensions, since the Einstein tensor identically vanishes according to the GR, the equation of motion for metric leads to $T_{\mu\nu}=0$. In our case, equation of motion for metric gives us the energy momentum tensor which identically vanishes. We note here that the trace of the energy momentum tensor reduces to the wave equation given in Eq.(\ref{eom-trace}) which can be obtained using a suitable linear combination of the components of the $(tt),(zz)$ of Eq.(\ref{eom-phi}). Consequently, we can only use either Eq.(\ref{trace}) or Eqs.(\ref{tt} \& \ref{zz}). In this work, we use only Eq.(\ref{trace}). Additionally, there is another off diagonal component of the energy momentum tensor equation, i.e, the off diagonal  $(tz)$ component. Note that following the above discussions, we now have only Eq.(\ref{trace}) along with Eq.(\ref{tz}). Any solution of Eq.(\ref{trace}) must be a solution of Eq.(\ref{tz}) as well. We just add these equations, i.e., equation of the trace Eq.(\ref{trace}) and Eq.(\ref{tz}). The resulting equation is a simple linear PDE presented in Eq.(\ref{eqphi}). It is clear to us here that any solution to the single linear PDE Eq.(\ref{eqphi}) can be used to solve the trace Eq.(\ref{trace}). Consequently, it is a solution to Eq.(\ref{tt} \& \ref{zz}) and also can be used to solve PDE Eq.(\ref{tz}). It is clear that there is a consistency between equations we used in our study. There is no any extra solution to the equations because we didn't differentiate any equation.

\subsection{Hyperbolic space}
In the first case, we consider space when $\Omega= \{z\in[0,\infty] \times t\in[-\infty , \infty]\}$. Using the standard Fourier transformation of the dilaton profile $\phi(t,z)$ as 
\begin{eqnarray}
\phi(t,z)=\int_{-\infty}^{\infty}\tilde{\phi}(\omega,z)e^{i\omega t}d\omega\,.\label{FT}
\end{eqnarray}
One can show that the Fourier amplitude function $\tilde{\phi}(\omega,z)$ satisfies the following simple ODE :
\begin{eqnarray}
\tilde{\phi}(\omega,z)-i\omega z(z\tilde{\phi}(\omega,z))'=0\,.\label{amp}
\end{eqnarray}
We can simply solve Eq.(\ref{amp}) to obtain an exact solution for the Fourier amplitude and find that
 \begin{eqnarray}
\tilde{\phi}(\omega,z)=\frac{c_1}{z}e^{\frac{i}{z\omega}}\,, 
\end{eqnarray}
where in general $c_1$ is a function of the frequency $\omega$. If we keep it as $c_1(\omega)$, then this Fourier amplitude needs to be fixed using an initial condition for the field, i.e, by adapting a regular profile $\phi(0,z)$. Note that the initial wave packet for the wave amplitude is still an open question since we do not know how the propagation initiated at the beginning. However, it is always possible to consider a monochromatic wave packet, i.e when $c_1(\omega)$  is sharply peaked around a central frequency $\omega_0$. In this approximation, we can use the saddle point approximation and basically take the Fourier amplitude $c_1(\omega)\approx c_1$ as a constant. Otherwise, the complete wave pattern needs an information about the initial wave packet which is not available as initial data here. During this work we always adapt the monochromatic approximation for wave amplitudes. Taking into account the above approximation for Fourier amplitude $c_1(\omega)$ and then inserting this Fourier amplitude into the Fourier integral transformation Eq.(\ref{FT}), we obtain
  \begin{eqnarray}
\phi(t,z)&=&\frac{c_1}{\sqrt{i\pi}
|zt| ^{3/2}}\Big((| t| +i t) \ker _1(\frac{2 \sqrt{| t| }}{\sqrt{i z}})+(t-i | t|)\text{kei}_1(\frac{2 \sqrt{| t| }}{\sqrt{-i z}})\nonumber\\&&+(t+i| t| ) (\ker_1(\frac{2 \sqrt{| t| }}{\sqrt{-i z}})+i \text{kei}_1(\frac{2 \sqrt{|
   t| }}{\sqrt{i z}}))\Big)\,,
\end{eqnarray}
 where ${\rm ker}_{1}$ and $\text{kei}_1$ are Kelvin functions. It is worth noting that the solution given above covers a whole hyperbolic spacetime. The AdS boundary value for the dilaton field can be simply obtained when setting $z=\epsilon\to 0$. Using the series expansion of the Fourier amplitude in the vicinity of $z=\epsilon$, we can evaluate the following inverse Fourier integral:
 \begin{eqnarray}
\phi(t,\epsilon)=\frac{c_1}{\epsilon}\int_{-\infty}^{\infty}e^{i(\omega t+\frac{1}{\omega \epsilon})}d\omega+\mathcal{O}(z-\epsilon)\,.
\end{eqnarray}
 It is not possible to analytically solve the above integral. Basically an integration gives us a lengthy expression of transcendental functions. There is a term of the integral which depends of the UV cutoff $\epsilon^{-1}$ and that term is given by $2\pi \delta(t)$. The leading terms can be expanded in series of orders of $\mathcal{O}(\epsilon^n),\,n>1$. Finally, the boundary value for a dilaton profile is approximately given by 
 \begin{eqnarray}
\phi_{bdy}(t)\approx \frac{c_1\sqrt{2\pi}}{\epsilon}\delta(t)+\mathcal{O}(\epsilon^n)\,.
\end{eqnarray}
In comparison to the general AdS/CFT program, see e.g. the discussion around Sec.4.6 of Ref.\cite{Maldacena:2003nj}, the boundary value for the field is scaled as $\epsilon^{-1}$. This implies that the renormalized field is $\phi_r\propto \delta (t)$ yielding the existence of a boundary shock field amplitude. In the language of the CFT, the conformal dimension is fixed as $\Delta_{-}=-1$ for a typical dilaton field on JT gravity with AdS$_2$ classical Euclidean background. 
 \subsection{Hyperbolic cusp geometry} 
In JT gravity, equations of motion provide pure hyperbolic (AdS) as a unique geometry for the gravity sector. The dual to pure AdS could be considered as a $T=0$ system with CFT or nearly-CFT symmetry. In the global coordinates of the AdS, in the absence of a black hole there is no way to introduce temperature to the system. There is a possibility to cut the AdS global patch and make it thermal. It is equivalent to make the time coordinate bounded instead of extending it to the entirely plane. This cut scenario is corresponding to the cusp geometry in hyperbolic geometries. Generally speaking, a cusp geometry as a cut of the AdS is defined as a geometry when the boundary is geodesic with cusps in the marked points. The geometry is appeared in the open string theory. From the mathematical point of view, a cusp geometry is considered as the moduli space of bordered Riemann surfaces. In our two-dimensional geometry set up, the cusp geometry corresponds to the metric with AdS boundary located at $z=0$ with marked points \cite{2010arXiv1002}. There were several studies about possible dualities as AdS$_2$/open string or AdS$_2$/CQM. For example, it was discovered that AdS$_2$ gravity is dual to conformally invariant mechanical system \cite{Cadoni:2000gm}. From the mathematical physics point of the view, it is interesting to study dilaton profile solutions for the cusp geometry as a  bounded region in the hyperbolic space.
 In this cusp geometry for which $ \Omega=\{z\in[0, \infty ]\times  t\in[0,1]\}$, it is convenient to expand the dilaton field in terms of an orthonormal basis of the functions as follows:
\begin{eqnarray}
\phi(t,z)=\frac{\psi_0(z)}{2}+\sum_{n=1}^{\infty}\Big(\varphi_n(z)\sin(n\pi t)+\psi_n(z)\cos(n\pi t)
\Big)\,.\label{cusp1}
\end{eqnarray}
By substituting the Fourier form given in Eq.(\ref{cusp1}) into the dilaton field equation (\ref{eqphi}), one can show that $\psi_0(z)\equiv 0$ and the Fourier amplitudes $\varphi_n(z),\psi_n(z)$ satisfy the following set of the operator equation for each mode with $n\geq 1$:
   \begin{eqnarray}
\hat{\mathcal{O}}_n(z)\{\varphi_n(z)\}=\psi_n(z),\ \ \hat{\mathcal{O}}_n(z)\{\psi_n(z)\}=-\varphi_n(z).
\label{cusp2}
\end{eqnarray}
Here the differential operator $\hat{\mathcal{O}}_n(z)$ is defined as
  \begin{eqnarray}
\hat{\mathcal{O}}_n(z)\equiv n\pi z(1+z\frac{d}{dz}),\ \ z\in[0, \infty ].
\label{cusp3}
\end{eqnarray}
Rewritten the above system of ODEs given in Eqs.(\ref{cusp2}) in Hamiltonian form, this equations form an infinite-dimensional dynamical system which can be written in matrix form as follows: 
\begin{equation}
  \begin{bmatrix}
    1 & \hat{\mathcal{O}}_n(z)\\
\hat{\mathcal{O}}_n(z) & -1
  \end{bmatrix}
  \begin{bmatrix}
   \varphi_n(z) \\
  \psi_n(z)
    \\
  \end{bmatrix} 
  =0\,.
\end{equation} 
It is worth noting that the authors of \cite{Gottschalk:2018kqt} reformulated the evolution of the quantum state as an infinite-dimensional dynamical system with mathematical features different from the standard theory of infinite-dimensional dynamical systems.

Using a change of the variables as $z\to u=\ln z$, we find that the differential operator transforms as 
\begin{eqnarray}
\hat{\mathcal{O}}_n(u)\equiv n\pi e^{u} (1+\frac{d}{du}),\ \ u\in[-\infty, \infty ].
\label{cusp4}
\end{eqnarray}
Here it is written in terms of a variable $u$ in which the conversion can be straightforwardly done. 

\subsubsection{Exact dilaton profile}
We can integrate out the ODEs given in Eqs.(\ref{cusp2}) by using the auxiliary operator $[\hat{\mathcal{O}}_n(u)]^2$. Using this, we can get an exact solution which can be solved from the resulting equation:
\begin{eqnarray}
\varphi_n(u)''+\varphi(u)'+\frac{e^{2u}}{n^2\pi^2}\varphi(u)=0\,.\label{cusp5}
\end{eqnarray}
Once an exact solution for $\varphi(u)$ is quantified, then the other dilaton Fourier amplitudes can be obtained using an operation
$\psi_n(u)=\hat{\mathcal{O}}_n(u)\varphi_n(u)$. We solve for an exact solution for Eq.(\ref{cusp5}) to yield
\begin{eqnarray}
&&\varphi_n(u)=\sqrt{\frac{n\pi}{2}}e^{-u}\big(2c_{1n}\cos(\frac{e^u}{n\pi})+c_{2n}\sin(\frac{e^u}{n\pi})
\big)\,,\label{cusp6}\\&&
\psi_n(u)=\sqrt{\frac{n\pi}{2}}e^{-u}\big(c_{2n}\cos(\frac{e^u}{n\pi})-2c_{1n}\sin(\frac{e^u}{n\pi})
\big)\,,\label{cusp7}
\end{eqnarray}
and the dilaton profile reduces to the following expression:
\begin{eqnarray}
\phi(t,u)=e^{-u}\sum_{n=1}^{\infty}\sqrt{\frac{n\pi}{2}}
\Big(c_{2n}\cos(\frac{e^u}{n\pi}-n \pi t)-2c_{1n}\sin(\frac{e^u}{n\pi}-n \pi t)
\Big),
\label{cusp8}
\end{eqnarray}
with $u\in (-\infty,\infty)$. Invoking the AdS boundary (bdy) value $\phi_{\rm bdy}(t)$ as an initial condition where $\phi_{\rm bdy}(t)=\phi(t,-\infty)$ and a UV cutoff as $\Lambda=\lim_{u\to -\infty} e^{-u}$, then the Fourier coefficients $c_{1n},c_{2n}$ are obtained
\begin{eqnarray}
&&c_{1n}=\Lambda^{-1}\sqrt{\frac{2}{n\pi}}\int_{0}^{1}\phi_{bdy}(t)\sin(n\pi t)\,,
\label{cusp9}\\&&
c_{2n}=2\Lambda^{-1}\sqrt{\frac{2}{n\pi}}\int_{0}^{1}\phi_{bdy}(t)\cos(n\pi t)\,.
\label{cusp10}
\end{eqnarray}
By specifying the initial bdy value of the fields, we can write the full dilaton profile using the Fourier series  (\ref{cusp8}). Some special classes of the bdy models are listed below: 
\begin{itemize}
\item Static boundary $\phi_{\rm bdy}(t)=\phi_{\rm bdy}(0)=\phi_{\rm bdy}$: the integrals can be simplified and we obtain
\begin{eqnarray}
&&c_{2n}=0,\ \ c_{1n}=\Lambda^{-1}\phi_{\rm bdy}\sqrt{\frac{2}{((2n+1)\pi)^3}},\, n=0,1,...\infty\,.\label{cusp11}
\end{eqnarray}
Notice that the dilaton profile in this case cannot be easily written in a closed form, but it is given by 
\begin{eqnarray}
\phi(t,u)=-2\Lambda^{-1}\phi_{\rm bdy}e^{-u}\sum_{n=0}^{\infty}\frac{\sin(\frac{e^u}{(2n+1) \pi}-(2n+1) \pi t)}{\pi (2n+1)}\,.
\label{cusp12}
\end{eqnarray}

\item Shock boundary regime $\phi_{\rm bdy}(t)=\phi_{c}\delta(t-t_c),\,0<t_c<1$: In this case, we have
\begin{eqnarray}
&&c_{1n}=\Lambda^{-1}\phi_c\sqrt{\frac{2}{n\pi}}\sin(n\pi t_c)\,,
\label{cusp13}\\&&
c_{2n}=2\Lambda^{-1}\phi_c\sqrt{\frac{2}{n\pi}}\cos(n\pi t_c)\,.
\label{cusp14}
\end{eqnarray}
We find that the full dilaton profile in this case has a closed form as
\begin{eqnarray}
\phi(t,u)=2\Lambda^{-1}\phi_ce^{-u}\sum_{n=1}^{\infty}\cos(\frac{e^u}{n\pi}-n \pi (t-t_c))\,.
\label{cusp15}
\end{eqnarray}

\end{itemize}

\subsection{Geometry with $\Omega=\{z\in [0,1]\times t\in [0,\infty)\}$}
We first assume the Euclidean time $t\in[0,\infty)$. One of the effective methods to solve it is to use a Laplace transform along a suitable boundary conditions for the domain $\Omega=[0,1]\times [0,\infty)$. A Laplace transform of Eq.(\ref{eqphi}) is given by 
\begin{eqnarray}
\phi(t,z)=\int_{0}^{\infty}\tilde{\phi}(s,z) e^{-st}dt.
\end{eqnarray}
Substituting it into Eq.(\ref{eqphi}), we obtain
\begin{eqnarray}
sz^2\frac{\partial}{\partial z}\tilde{\phi}(s,z)+(sz-1)\tilde{\phi}(s,z)+(-z\phi(0,z)-z^2\phi'(0,z))=0 \label{tildephi}.
\end{eqnarray}
Using the standard procedure, we can find exact solutions for the field amplitude $\tilde{\phi}(s,z) $ in the Laplace plane $s$ and then by performing an inverse Laplace transform one can obtain the general dilaton profile ${\phi}(t,z)$. We need initial condition (IC) to solve it. In particular the following boundary conditions are introduced:
\begin{eqnarray}
&&
\phi(t,z)=
\left\{
	\begin{array}{ll}
		0 & \mbox{if } t\to+\infty\\
		f(z)& \mbox{if } t=0.
	\end{array}
\right.
\end{eqnarray}
As we see, the $\tilde{\phi}(s,z)$ depends on the IC function, $f(z)$. An exact solution for $\tilde{\phi}(s,z)$ is obtained as follows:
\begin{eqnarray}
\tilde{\phi}(s,z)=\frac{c_1}{z}e^{-\frac{1}{sz}}+\frac{f(z)}{s}-\frac{f(1)}{sz}e^{-\frac{1-z}{sz}}-\frac{e^{-\frac{1}{sz}}}{s^2 z}\int_{z}^{1}\frac{f(w)e^{\frac{1}{sw}}}{w}dw.
\label{tildephi-sol}
\end{eqnarray}
Note that $z\in[0,1]\leq 1$. Again, we have used the monochromatic approximation and also assumed that $c_1$ is just a constant. In particular, we can check that the value of the field amplitude $\tilde{\phi}(s,z)$ at the point $z=1$ is given by
\begin{eqnarray}
\tilde{\phi}(s,1)=c_1e^{-\frac{1}{s}}.
\end{eqnarray}
Hence one can use inverse Laplace transform to obtain $\phi(t,1)=\frac{J_1(2\sqrt{t})}{\sqrt{t}}$. Now, the complete solution using inverse Laplace transform of the above expression can be obtained and we find
\begin{eqnarray}\label{phi-tz-final}
\phi(t,z)&=&f(0)-\frac{f(1)}{z}J_0(2\sqrt{\frac{t}{z}})-\frac{c_1}{(tz)^{3/2}}J_1(2\sqrt{\frac{t}{z}})\nonumber\\&&+\frac{c_1\delta (t)}{z}
-\sqrt{t}\int_{z}^{1}\frac{dw f(w)J_1(2\sqrt{\frac{t(w-z)}{wz}})}{\sqrt{-wz^2+w^2z}}\,,
\end{eqnarray}
where $f(0),f(1),c_1 $ are arbitrary parameters. 
\begin{itemize}
\item Evaluation of $\phi(z,t)$ for a class of critical $f(z)$:
We can evaluate the integral appeared in the exact dilaton profile given in Eq.(\ref{phi-tz-final}) for the following simple but physically significant cases:
\begin{itemize}
    \item If $f(w)=\delta(w-w_0)$ is a localized dilaton profile at the initial time, the exact profile takes the following form:
\begin{eqnarray}
\phi(t,z)&=&f(0)-\frac{f(1)}{z}J_0(2\sqrt{\frac{t}{z}})-\frac{c_1}{(tz)^{3/2}}J_1(2\sqrt{\frac{t}{z}})\nonumber\\&&+\frac{c_1\delta (t)}{z}-
		\frac{t }{w_0
  z}\theta (1-w_0,w_0) \, _0 F_1\big(2;t
   \big(\frac{1}{w_0}-\frac{1}{z}\big)\big), 
\end{eqnarray}
where $w_0\in \mathbb{R}$ and $\theta (x_1,x_2)$ defines the multivariables Heaviside theta function, which is equal $1$ only if all of the arguments $x_i$ are positive and $_0 F_1\big(2;t
   \big(\frac{1}{w_0}-\frac{1}{z}\big)\big)$ is the generalized hypergeometric function. 

    \item If $f(w)=\theta(w-w_0)$ is a step function as a shock wave at initial time, 
\begin{eqnarray}
\phi(t,z)=f(0)-\frac{f(1)}{z}J_0(2\sqrt{\frac{t}{z}})-\frac{c_1}{(tz)^{3/2}}J_1(2\sqrt{\frac{t}{z}})+\frac{c_1\delta (t)}{z}\,,
\end{eqnarray}
with $\Re(w_0)>1,\,\Im(w_0)=0$ and $J_{i}(x)$ are Bessel functions. 
\end{itemize}

\item Dilaton on AdS boundary: In Ref. \cite{Maldacena:2016upp}, it is assumed that $\phi_{bdy}(t)\equiv \phi(t,z=0)$ is constant, but here we keep it as an arbitrary function of the Euclidean time $t$. In the limiting case, $z\to\epsilon$, we find $\phi(t,z\to\epsilon)=\phi_{\rm bdy}(t)$. We can simply do series expansion up to any order of $\epsilon$ as well as study the $\phi_{bdy}(t,z\to\epsilon)$ for both regimes, i.e. initial time $t=0$ and late time $t=\infty$, using the exact profile given in Eq.(\ref{phi-tz-final}).  Furthermore, we can quantify the boundary value via the Laplace amplitude $\tilde{\phi}(s,z)$. Basically, we have
\begin{eqnarray}
&&\tilde{\phi}(s,z\to \epsilon)=\frac{c_1}{\epsilon}e^{-\frac{1}{s\epsilon}}+\frac{f(0)}{s}-\frac{f(1)}{s\epsilon}e^{-\frac{1}{s\epsilon}}-\frac{e^{-\frac{1}{s\epsilon}}}{s^2 \epsilon}\int_{0}^{1}\frac{f(w)e^{\frac{1}{sw}}}{w}dw.\label{phi-zs}
\end{eqnarray}
By performing series expansion, we can show that the boundary value of the dilaton field is given approximately by the following expression:
\begin{eqnarray}
\phi_{\rm bdy}\approx f(0)-\frac{\beta }{\sqrt[4]{\epsilon }}-\frac{f(1)}{\epsilon
   ^{3/4}}+\frac{\zeta }{\epsilon ^{5/4}}+\frac{1}{\epsilon }\,.
\end{eqnarray}

\item Initial value of the field on the AdS boundary: At initial time $t=0$, we have to take the limit of the expression Eq.(\ref{phi-zs}) and keep $\epsilon\ll 1$ \& finite. Then we find 
\begin{eqnarray}
\phi_{\rm bdy}(0,z\to\epsilon)=f(0)-\frac{f(1)}{\epsilon}+\frac{c_1\delta (0)}{\epsilon}.
\end{eqnarray}
In the CFT language, the conformal dimension here is $\Delta_{-}=-1$.
\item Late time value of the field on the AdS boundary: For a late time, when $t\to \infty$, we have
\begin{eqnarray}
\phi_{\rm bdy}(\infty,z\to\epsilon)\approx f(0)
-\sqrt[4]{t}\int_{\epsilon}^{1}\frac{dw f(w)\cos(2\sqrt{t(\frac{1}{\epsilon}-\frac{1}{w})}-\frac{3\pi}{4})}{w}\,.
\end{eqnarray}
We can show that for all $w\in [\epsilon,1]$ it provides that $|\cos(2\sqrt{t(\frac{1}{\epsilon}-\frac{1}{w})}-\frac{3\pi}{4})|\leq 1$. Consequently, we have
\begin{eqnarray}
|\phi_{bdy}(\infty,z\to\epsilon)|\leq  |f(0)|
+\sqrt[4]{t}\int_{\epsilon}^{1} |f(w)| d|w| \,.
\end{eqnarray}
If $f(w)\propto w^{n},n\neq -1$, we then obtain
\begin{eqnarray}
|\phi_{\rm bdy}(\infty,z\to\epsilon)|\leq  |f(0)|
+\frac{\sqrt[4]{t}}{n+1}\,.
\end{eqnarray}
However, for the case with $f(w)\propto \frac{1}{w}$, then we find
\begin{eqnarray}
\lim_{t\to \infty}  \frac{|f(0)|-|\phi_{\rm bdy}(t,\epsilon)|}{\log  \epsilon \sqrt[4]{t} } \geq 1\,.
\end{eqnarray}
Note that the last case is a generic case of the AdS in two dimensions and it doesn't obey the power-law behavior for boundary value of the $\phi$ in other dimensions (see for example  the discussion around Sec.4.6 of Ref.\cite{Maldacena:2003nj} for boundary value of a generic scalar field in dimensions different from two).

 \end{itemize}
\section{Deformed Jackiw-Teitelboim gravity}
\label{dJT}
We next consider deformed JT (dJT) gravity investigated recently in Refs \cite{Maxfield:2020ale},\cite{Witten:2020ert}. The model is described easily by including a potential term to the original JT action. It takes the form
\begin{eqnarray}
S=-\frac{1}{2}\int d^2x\sqrt{g}\Big(\phi(R+2)+W(\phi)\Big)\,.\label{S2}
\end{eqnarray}
We take $W$ as a general function of $\phi$. The field equations derived from action Eq.(\ref{S2}) are simply the vanishing Einstein tensor and a type of the Klein-Gordon equation for the dilaton field $\phi$:
\begin{eqnarray}
&&R+2+W'(\phi)=0\label{eom-g2},\\&& \nabla_{\mu}\nabla_{\nu}\phi-g_{\mu\nu}\nabla_{\alpha}\nabla^{\alpha}\phi+\Big(\phi+\frac{W(\phi)}{2}\Big)g_{\mu\nu}=0\,.\label{eom-phi2}
\end{eqnarray}
The pure AdS solution exist as an exact solution if and only if the potential function satisfies the following constraints:
\begin{eqnarray}
W'(\phi_c)=0,\,\,W(\phi_c)=-2\phi_c\,,\label{W,phi_c}
\end{eqnarray}
with a uniform dilaton profile $\phi=\phi_c$. For example with the potential function as $W(\phi)=-2\Lambda$ (an effective negative cosmological constant term), we have an exact AdS for a uniform profile $\phi=\phi_c$. If $\phi\neq \phi_c$, there is also a possibility to still have pure AdS if and only if $W(\phi)\equiv \mbox{const.}$, which we need to investigate very carefully in detail in the next section. Note here that the method to unify all of the equations of motion used in the section \ref{dJT} is similar to those of the pure JT gravity discussed in Sec.\ref{sec3}.
\subsection{Exact pure AdS in dJT for non uniform dilaton $\phi\neq \phi_c$}
For a constant scalar profile, there is a trivial class of the AdS solutions. But if $\phi\neq \mbox{const.}$, to have AdS we need $W(\phi)$ to satisfy the following relations:
\begin{eqnarray}
&&W'(\phi)=0 \label{eq1-deformation-pure AdS},\\&& \nabla_{\mu}\nabla_{\nu}\phi-g_{\mu\nu}\nabla_{\alpha}\nabla^{\alpha}\phi+\Big(\phi+\frac{W(\phi)}{2}\Big)g_{\mu\nu}=0\label{eq2-deformation-pure AdS}.
\end{eqnarray}
We adapt once the Euclidean metric Eq.(\ref{metric1}) and the non-vanishing components of the Einstein equation read 
\begin{eqnarray}
&&(tt): {\phi}''-\frac{\phi}{z^2}-\frac{W(\phi)}{2z^2}=0\,,\label{tt2}\\&&
(tz): \dot{\phi}'+\frac{\dot{\phi}}{z}-\ddot{\phi}-\phi''+\frac{\phi}{z^2}+\frac{W(\phi)}{2z^2}=0\,,\label{tz2}\\&&
(zz): \ddot{\phi}-\frac{\phi}{z^2}-\frac{W(\phi)}{2z^2}=0\,,\label{zz2}
\end{eqnarray}
with $W'(\phi)=0$. From this constraint, we obtain, $W(\phi)=W_0$, similar to the pure JT. In this case, the trace equation reads $\nabla_{\mu}\nabla^{\mu}\phi-(2\phi+W_0)=0$ which can be simply derived using the above EoMs (\ref{tt2} and \ref{zz2}). By adding trace equation to the off diagonal equation 
(\ref{tz2}), we end up with the following PDE:
\begin{eqnarray}
&&z(z\dot{\phi})'-\phi-\frac{W_0}{2}=0.
\label{eq1}
\end{eqnarray}
This is a modified version of the JT equation given in Eq.(\ref{eqphi}). Note that any solution for the above PDE (\ref{eq1}) solves all the EoMs (\ref{tt2}-\ref{zz2}).  We can investigate all types of the solutions as those we already discussed in the pure JT gravity. We are interested in the Euclidean hyperbolic coverage of the AdS where one can use Fourier transformation. Therefore the Fourier amplitude can be obtained as
\begin{eqnarray}
&&\tilde{\phi}(\omega,z)-i\omega z(z\tilde{\phi}(\omega,z))'+\frac{W_0}{2}\delta(\omega)=0\,.
\end{eqnarray}
An exact solution for the Fourier amplitude cam be determined to obtain
 \begin{eqnarray}
\tilde{\phi}(\omega,z)=\frac{c_1}{z}e^{\frac{i}{z\omega}} +\frac{iW_0e^{\frac{i}{z\omega}}}{2\omega z}\delta(\omega)Ei(-\frac{i}{\omega z})\,,
\end{eqnarray}
where $c_1$ is an integration constant and $Ei(z)=-\int_{-z}^{\infty}\frac{e^{-t}}{t}dt$ gives the exponential integral function. By inserting this Fourier amplitude into the Fourier integral transformation Eq.(\ref{FT}), we find
  \begin{eqnarray}
\phi(t,z)&=&\frac{c_1}{\sqrt{i\pi}
   | zt| ^{3/2}}\Big((| t| +i t) \ker _1(\frac{2 \sqrt{| t| }}{\sqrt{i z}})+(t-i | t| )
   \text{kei}_1(\frac{2 \sqrt{| t| }}{\sqrt{-i z}})\nonumber\\&&\quad\quad\quad\quad\quad+(t+i | t| ) (\ker
   _1(\frac{2 \sqrt{| t| }}{\sqrt{-i z}})+i \text{kei}_1(\frac{2 \sqrt{|
   t| }}{\sqrt{i z}}))
   \Big)\nonumber\\&&+\frac{1}{2\pi}\int_{-\infty}^{\infty}\frac{iW_0e^{\frac{i}{z\omega}-\omega t}}{2\omega z}\delta(\omega)Ei(-\frac{i}{\omega z})d\omega\,.
\end{eqnarray}
The integral in the last line can be computed using the counter integral by finding the residue for the integrand at pole $\omega=0$. The result contains a divergence term $\delta(0)$ .

\subsection{Non AdS solutions for the dJT}
Consider the system of the EOMs given in the Eqs.(\ref{eom-g2},\ref{eom-phi2}). We assume that there may be some non AdS black hole solutions (e.g. solitons, topological defects, and so forth) where $R\neq -2$.  In this case we find
\begin{eqnarray}
&&R+2+U'(\phi)=0\,,\label{eom-g3}\\&& \nabla_{\mu}\nabla_{\nu}\phi-g_{\mu\nu}\nabla_{\alpha}\nabla^{\alpha}\phi+(\phi+\frac{U(\phi)}{2})g_{\mu\nu}=0\,.\label{eom-phi3}
\end{eqnarray}
Here we can adapt a kind of light-cone coordinate system $(u,v) $ in the spacetime, but we have to first show that such metric is accessible. To do this, we start with a simple general Euclidean time independent (stationary)  metric in the Poincar\'e coordinates $(t,z)$:
\begin{eqnarray}
ds^2=f(z)(dt+\frac{h(z)}{f(z)}dz)^2-\frac{h^2(z)+k(z)f(z)}{f(z)}dz^2\,.
\end{eqnarray}
One can define $\xi=t+\int \frac{h(z)}{f(z)}dz$ and therefore the metric reduces to
\begin{eqnarray}
ds^2=f(z)d\xi^2-\frac{h^2(z)+k(z)f(z)}{f(z)}dz^2\,.
\end{eqnarray}
Now if one define the light-cone coordinate system as
\begin{eqnarray}
&&u=\xi-\int \frac{\sqrt{h^2(z)+k(z)f(z)}}{f(z)}dz\,,\\&&
v=\xi+\int \frac{\sqrt{h^2(z)+k(z)f(z)}}{f(z)}dz\,,
\end{eqnarray}
we can finally write a general metric as
\begin{eqnarray}
ds^2=f(u,v)dudv,\ \ f(u,v)=f(z(u,v))\,.
\label{metric3}
\end{eqnarray}
 Rewriting Eqs.(\ref{eom-g3},\,\ref{eom-phi3}) and using the metric (\ref{metric3}), we then obtain
\begin{eqnarray}
&&\partial_u\partial_v\log f(u,v)+(1+\frac{U'(\phi)}{2})\frac{f(u,v)}{2}=0\,,
\label{eom-g4}
\end{eqnarray}
as well as
\begin{eqnarray}
&&(uu): \partial^2_u\phi(u,v)-\partial_u\log f(u,v)\partial_u\phi(u,v)=0\,,
\label{uu4}\\&&
(uv): (\phi+\frac{U(\phi)}{2})f(u,v)-\partial_u\partial_v\phi(u,v)=0\,,
\label{uv4}\\&&
(vv): \partial^2_v\phi(u,v)-\partial_v\log f(u,v)\partial_v\phi(u,v)=0\,.
\label{vv4}
\end{eqnarray}
We can integrate out Eq.(\ref{vv4}) to yield
\begin{eqnarray}
\partial_v\phi(u,v) =\alpha (u)f(u,v)\,,
\label{vv44}.
\end{eqnarray}
where $\alpha (u)$ denotes an arbitrary function on $u$. Similarly using Eq.(\ref{uu4}), we obtain
\begin{eqnarray}
\partial_u\phi(u,v) =\beta(v)f(u,v)
\label{uu44}.
\end{eqnarray}
A consistency relation for Eqs(\ref{vv44},\,\ref{uu44}) provides us with
\begin{eqnarray}
\partial_u(\alpha (u)f(u,v)) =\partial_v(\beta(v)f(u,v))
\label{con1}.
\end{eqnarray}
Fortunately it can be simplified to obtain the following PDE for $f(u,v)$:
\begin{eqnarray}
\beta(v)\partial_v \log f(u,v)-\alpha(u)\partial_u \log f(u,v)=\alpha' (u)-\beta'(v)
\label{con2}.
\end{eqnarray}
If we change the variables from $v\to V= \int\beta^{-1}(v)dv,\,u\to U= \int\alpha ^{-1}(u)du $, we can simplify PDE given in Eq.(\ref{con2}) as follows:
\begin{eqnarray}
\partial_V \log f(U,V)-\partial_U \log f(U,V)=\partial_U \log \alpha(U)-\partial_V \log \beta(V)
\label{con3}.
\end{eqnarray}
Notice that this is an inhomogenous first-order PDE and the solution can be easily found in the following cases:

\begin{itemize}
    \item
Case I: $\alpha(U)={U^{M+1}},\,\beta(V)={V^{N+1}},\,M,N\neq -2$: We find
\begin{eqnarray}
f(U,V)=\exp\big(P(U+V)-(1+M)\log U-(1+N)\log V\big)\,,
\label{con4}
\end{eqnarray}
where $P$ is an arbitrary function. Now we can integrate out the PDE given in Eq.(\ref{uv4}). We need to transform back to the coordinates from $U,V\to u,v $ up to some arbitrary integration constants to obtain
\begin{eqnarray}
&& V=-\frac{v^{-N}}{N},\,\, U=-\frac{u^{-M}}{M}\,.
\end{eqnarray}
Therefore the metric function changes to the following equivalent form in the original null coordinates system:
\begin{eqnarray}
f(u,v)=\exp\Big(P(\frac{u^{-M}}{M}+\frac{v^{-N}}{N}
)+\frac{u^{-M}(1+M)}{M}+\frac{v^{-N}(1+N)}{N}\Big)\,.
\label{con51}
\end{eqnarray}
Note that the solution given in Eq.(\ref{con51}) is a family of the exact solutions for the theory and it is different from the AdS. It is worth mentioning that a soliton solution for the PDE given in Eq.(\ref{uv4}) provides us an exact profile for the dilaton field. Basically one can solve Eq.(\ref{uv4}) along the exact metric presented in Eq.(\ref{con51}) to yield
\begin{eqnarray}
&& \phi(u,v)=-\frac{v^{-N(N+1)}}{(-N)^{N+1}}\int\partial {u}f(u,v)
\label{con44444}.
\end{eqnarray}

\item 
Case II: $ \alpha(U)={U^{-1}},\beta(V)={V^{-1}
}$: We have
\begin{eqnarray}
f(U,V)=Q(U+V)+\log(-UV)\,.
\label{con4}
\end{eqnarray}
We can transform back to the original coordinates $(u,v)$ and use $u=\sqrt{2|U|},v=\sqrt{2|V|}$. As a result, the metric function reads
\begin{eqnarray}
f(u,v)=Q(\frac{u^2+v^2}{2})+\frac{1}{2}\log(\frac{uv}{2})
\label{con4}.
\end{eqnarray}
In this simple case, the dilaton profile can be obtained by integrating the following (exact) Pfaffian form
\begin{eqnarray}
d\phi(u,v)=f(u,v)(\frac{du}{u}+\frac{dv}{v})\,,
\label{con444}
\end{eqnarray}
which can be integrated easily to give
\begin{eqnarray}
&& \phi(u,v)=-\frac{1}{v}\int\partial {u}f(u,v)
\label{con4444}.
\end{eqnarray}

\end{itemize}


\section{Conclusions}

In this work, we investigated exact solutions for both JT and deformed JT gravity recently proposed by ~Maxfield-~Turiaci and \,Witten. We revisited the Jackiw-Teitelboim theory and examined its exact Euclidean solutions using all the non-zero components of the dilatonic equations of motion using proper integral transformation over Euclidean time coordinate. More precisely, we solved the system of field equations to find exact solutions for hyperbolic coverage, cusp geometry and another compact sector of the AdS$_2$ spacetime manifold. Then we considered its deformation called dJT gravity and figured out the solutions for both AdS and non AdS cases. The deformation can be simply done by adding a potential term $U(\phi)$ to the action of the JT gravity. 

Regarding our present work, it is possible to extend this study to account of higher-order corrections to JT gravity from the phenomenological point of the view to explore dual boundary theory. Yet, in a duality point of view, it was discovered that  AdS$_2$ gravity is dual to conformally invariant mechanical system \cite{Cadoni:2000gm}. From this, it is interesting to study dilaton profile solutions for the cusp geometry as a  bounded region in the hyperbolic space. Additionally, some other physical deformations may be worth investigating, see, e.g. $T{\bar T}$ deformations \cite{Dubovsky:2017cnj,Conti:2018tca,Ishii:2019uwk,Kim:2020eqn}. We will leave these interesting topics for further investigations.

\subsection*{Acknowledgments}
The work for D. Momeni is supported by the Internal Grant (IG/SCI/PHYS/20/07) provided by Sultan Qaboos University. We thank "Wolfram Research " for providing an "Extended Mathematica Evaluation License " used for computations in this work. We thank D. Stanford for comments and discussions about Ref. \cite{Maldacena:2016upp} and J. Maldacena for giving references about boundary in AdS$_2$ during a discussion section in (IGST) 2020. P. Channuie acknowledged the Mid-Cereer Research Grant 2020 from National Research Council of Thailand under a contract No. NFS6400117.

\end{document}